\PassOptionsToPackage{usenames,dvipsnames}{xcolor}
\documentclass[a4paper,11pt]{article}
\usepackage[version=3]{mhchem}
\usepackage{pos}
\usepackage{amsmath,pdftexcmds}
\usepackage{booktabs}
\usepackage{multirow}
\usepackage{tabularx}
\usepackage{amsfonts}
\usepackage{mathtools}
\usepackage{amsbsy}
\usepackage[T1]{fontenc}
\usepackage{dsfont}
\usepackage{hyperref}
\usepackage{xfrac}
\usepackage{pgfplots}
\usepackage{pgfplotstable}
\usepackage{tikz}
\usepackage{tikz-3dplot}
\usetikzlibrary{trees}
\usetikzlibrary{decorations.markings}
\usetikzlibrary{calc}
\usetikzlibrary{plotmarks}
\usetikzlibrary{fpu,arrows} 
\usepgfplotslibrary{fillbetween}

\definecolor{DarkMidnightBlue}{rgb}{0.0, 0.04, 0.14}
\definecolor{Bittersweet}{rgb}{1.0, 0.44, 0.37}
\definecolor{Burgundy}{rgb}{0.5, 0.0, 0.13}
\definecolor{Caribbeangreen}{rgb}{0.0, 0.8, 0.6}
\definecolor{Lilla}{rgb}{0.71, 0.4, 0.82}
\definecolor{Hotmagenta}{rgb}{1.0, 0.11, 0.81}
\definecolor{Tangerine}{rgb}{0.95, 0.52, 0.0}

\renewcommand{\logo}{\relax}

\hypersetup{hidelinks,
backref=true,
pagebackref=true,
hyperindex=true,
breaklinks=true,
colorlinks=true,
linkcolor=blue, 
citecolor=blue, 
urlcolor=blue,
bookmarks=true,
bookmarksopen=false,
pdftitle={Title},
pdfauthor={Author}}

\DeclareFontFamily{U}{cbgreek}{}
\DeclareFontShape{U}{cbgreek}{m}{n}{
        <-6>    grmn0500
        <6-7>   grmn0600
        <7-8>   grmn0700
        <8-9>   grmn0800
        <9-10>  grmn0900
        <10-12> grmn1000
        <12-17> grmn1200
        <17->   grmn1728
      }{}
\DeclareFontShape{U}{cbgreek}{bx}{n}{
        <-6>    grxn0500
        <6-7>   grxn0600
        <7-8>   grxn0700
        <8-9>   grxn0800
        <9-10>  grxn0900
        <10-12> grxn1000
        <12-17> grxn1200
        <17->   grxn1728
      }{}

\DeclareRobustCommand{\digamma}{%
  \text{\usefont{U}{cbgreek}{\normalorbold}{n}\symbol{147}}%
}

\makeatletter
\newcommand{\normalorbold}{%
  \ifnum\pdf@strcmp{\math@version}{bold}=\z@ bx\else m\fi
}
\makeatother

\title{Nuclear lattice effective field theory as a testing ground for $\alpha$-cluster structures in \ce{^{24}Mg}}
\ShortTitle{NLEFT as a testing ground for $\alpha$-cluster structures in \ce{^{24}Mg}}

\author*[a]{Gianluca Stellin}
\author[b]{Serdar Elhatisari}
\author[c]{Timo A. L\"ahde}
\author[d,e]{Shihang Shen}

\affiliation[a]{IJCLab, In2p3, CNRS and \textit{Université Paris-Saclay}, 91405 Orsay, France}
\affiliation[b]{\textit{Faculty of Natural Sciences and Engineering}, \textit{Gaziantep Islam Science and Technology University}, Gaziantep 27010, Turkey}
\affiliation[c]{\textit{Institute for Advanced Simulation (IAS-4)}, 
\textit{Forschungszentrum J\"ulich, D-52425 J\"ulich, Germany}}
\affiliation[d]{\textit{Peng Huanwu Collaborative Center for Research and Education} \& \textit{International Institute for Interdisciplinary and Frontiers}, \textit{Beihang University}, 100191 Beijing, China}
\affiliation[e]{\textit{School of Physics}, \textit{Beihang University}, 102206 Beijing, China}

\emailAdd{gianluca.stellin@ijclab.in2p3.fr}

\abstract{The framework of \textit{nuclear lattice effective field theory} (NLEFT) is applied to $^{24}\mathrm{Mg}$, with the perspective of obtaining a model-independent density map of the geometry of a sample of excited states of the nucleus. The Hamiltonian incorporates Wigner-SU(4)-symmetric nuclear forces as well as the Coulomb interaction. The coupling constants of the spin-isospin symmetric nucleon-nucleon potentials have been adjusted in order to reproduce the experimental ground-state (g.s.) energy of $^{24}\mathrm{Mg}$ as well as the experimental \textit{Tjon ratio} between the binding energies of \ce{^3H} and \ce{^4He}. The ensuing parameter set turns out to be capable of capturing the experimental trend of the binding energy per nucleon, reproducing simultaneously within 1\% deviation the measured values for \ce{^{18}F}, \ce{^{22}Na}, \ce{^{26}Al}, \ce{^{28}Si}, \ce{^{30}P} and \ce{^{32}S}. Considerations based on the convergence rate of Euclidean-time extrapolations for the two lowest energy eigenvalues highlight the dual nature of the $0_1^+$ and $2_1^+$ states, of hybrid mean-field and $\alpha$-cluster type. For the latter, triaxial $\alpha$-cluster configurations seem to be favoured over the axially-symmetric ones, whereas oblate superdeformation might characterize a rotational band at $20$~MeV excitation energy. 
}

\FullConference{
\begin{center}
\begin{minipage}{0.45\columnwidth} 
\includegraphics[width=1.0\columnwidth]{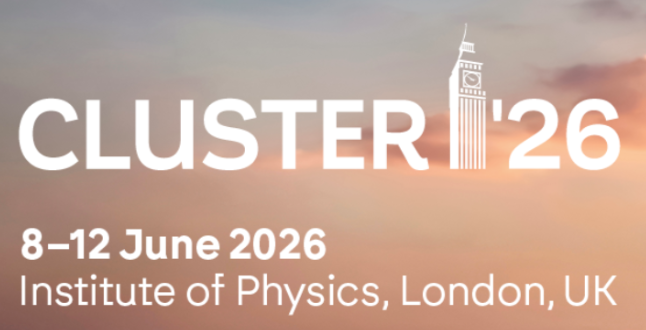} 
\end{minipage}\hspace{2mm}
\begin{minipage}{0.45\columnwidth}
\textbf{$12^{\mathrm{th}}$ International Conference on Clustering}\\
\textbf{Aspects of Nuclear Structure and Dynamics}, \\
June 8 - 12, 2026, \\
Institute of Physics,\\
37 Caledonian Road,\\
Islington, London N1 9BU, \\
United Kingdom
\end{minipage}
\end{center}
}

\begin{document}

\maketitle

\section{Preamble}

\noindent Since the discovery of the $\alpha$ particle, the structure of even-even self-conjugate nuclei is a subject of major interest in nuclear theory, as \ce{^4He} nuclei represent favoured cluster configurations \cite{Wef37,Whe37}. 

More generally, the observed nuclear spectra are the result of the interplay between indepen\-dent-particle effects in a mean field and spatial few-body correlations between the constituents \cite{KaE19}. While the former give rise to shell-structure features, the latter are responsible of clustering phenomena. Mainly in even-even self-conjugate nuclei, such correlations lead to the partial or complete partitioning of nuclear matter into $\alpha$-particle groupings. In recent literature, this twofold nature of the nuclear phenomenology is referred to as \textit{duality} \cite{Ohk22}-\cite{AZZ26}. 

The appearance of clusters affects the intrinsic structure of the nucleus, which acquires a \textit{molecular} shape \cite{Whe37,Gam30,HaT38}.  As a result, molecular or \textit{exotic} discrete symmetries \cite{Car97,DeD24,BKB25} add on top of continuous symmetries. Probably, the most striking example is represented by \ce{^{12}C}. Its ground state of is a mean-field state dominated by the $(0p_{3/2})^4$ closed-subshell configuration with significant overlap with $3\alpha$-cluster wavefunctions \cite{KaE19} with $\mathcal{D}_{3h}$ symmetry \cite{BiI02,KiT24}, indicating that $\alpha$-clusters are partially formed even in the lowest $0^+$ state. 

The most abundant isotope of magnesium, \ce{^{24}Mg}, makes no exception in this sense and has been recently investigated by means of a macroscopic $\alpha$-cluster model \cite{StS25,StS26}. There, the relevance of the point group $\mathcal{D}_{4h}$ \cite{Bou62,HaD66} for the prediction of spectrum and electromagnetic properties of the $^{24}\mathrm{Mg}$ nucleus has been discussed in the framework of the \textit{geometric $\alpha$-cluster model} (G$\alpha$CM) at leading order \cite{StS25,StS26}. The latter represents a macroscopic approach \cite{Ste15,SFV16} wherein nuclear excitations are described in terms of rotations and vibrations of $^4\mathrm{He}$ clusters about their equilibrium positions.

In Ref.~\cite{StS25}, it is found that reduced intraband $E2$ transition probabilities are better predicted by the square-bipyramidal rather than the bitetrahedral equilibrium configuration with $\mathcal{D}_{2h}$ symmetry \cite{HWD71}. Moreover, the identification of the candidates for the $9$ expected rotational bands corresponding to normal modes of vibration with a single quantum of excitation \cite{StS25,StS26} corroborates the $\mathcal{D}_{4h}$-symmetric \textit{ansatz}. Conversely, comparison between the experimental transition form factors and the LO G$\alpha$CM counterparts seem to favour $\mathcal{D}_{2h}$ symmetry, corresponding to either a bitetrahedron or a square bipyramid with the planar $\alpha$-clusters rotated by $\approx 30^{\circ}$ about an axis orthogonal to the main symmetry axis \cite{StS25} (cf. Fig.~1 of Ref.~\cite{SEL26}).

On the wave of the recent investigation on the $\alpha$-cluster structure of $^{12}\mathrm{C}$ \cite{SEL23} as well as earlier works \cite{EKL12,EKL14}, we try to crack the conundrum by applying the framework of NLEFT \cite{KBL18}-\cite{LLE19} to $^{24}\mathrm{Mg}$ \cite{SEL26}. The objectives include the construction of a model-independent density map of the geometry of the $J^{\pi} = 2_1^+$, $2_2^+$, $4_1^+$ and $3_1^+$ excitations, as well as the search for Hoyle-state analogues. An essential element of this study is represented by the \textit{pinhole} algorithm $^{12}\mathrm{C}$ \cite{SEL23}, which permits to obtain the nuclear density distribution associated with a given $A$-body state \cite{LLE19,YKL23}. Observables such as charge radii, form factors as well as $\gamma$-transition strengths are within the reach of the present code. 

\section{Theoretical framework}\label{sec:theoretical_framework}

NLEFT is an \textit{ab-initio} quantum Monte-Carlo approach (QMC) based on realistic nucleon-nucleon interactions, which stem from $\chi$EFT \cite{LaM19}. The Hamiltonian, $H = T+V$, exploited for the validation of key results for \ce{^{12}C} in Ref.~\cite{SEL23}, implements the $\chi$EFT forces up to N$^2$LO \cite{Lee09,MaS16,Heb21}. 

However, for the results illustrated here a Wigner-SU(4)-symmetric Hamiltonian with Coulomb interaction \cite{SEL23,LLE19} has been adopted. The spin-isospin-symmetric potentials consist of the two-body and three-body contact interactions in Refs.~\cite{SEL23,LLE19} with coupling constants $C_2$ and $C_3$ respectively  (cf. Tab.~\ref{tab:coupling-constants}). Their values are the result of new parameter fits.

Since most isospin-dependent forces tend to cancel in symmetric nuclear matter, whereas spin-dependent interactions are damped when summing over all possible nucleonic spin configurations \cite{LLE19}, Wigner-SU(4)-symmetric potentials prove to capture the salient features of the realistic nuclear forces in this context \cite{SEL23,LLE19}. At the same time, spin-isospin symmetric nuclear forces are capable of taming the \textit{sign problem} in NLEFT simulations \cite{LaM19,EBM24}. 

As a QMC method, the matrix elements of quantum operators linked to physical observables are determined by means of stochastic processes, regulated by the \textit{Metropolis-Hastings algorithm} \cite{Lee09,MRR53}. In this case, the Markov process enters Schr\"odinger's time-evolution operators, invoked for the convergence of trial $A$-body wavefunctions to exact eigenstates of the Hamiltonian operator, $H$. This coincides with the workflow of the \textit{adiabatic projection method} \cite{LaM19}, in which the time-evolution process is carried out in Euclidean time, $\tau \equiv it$ and the Monte-Carlo updates follow the \textit{shuttle} algorithm \cite{LLE19}.

\begin{table*}[htb!]
\begin{small}
\begingroup
\renewcommand{\arraystretch}{1.40}
\begin{center}
\begin{tabular}{c|c|c}
\toprule
\textsc{Method} & $C_2$ [$10^{-7}~(\hbar\mathrm{c})^3~\mathrm{MeV}^{-2}$] & $C_3$ [$10^{-14}~(\hbar\mathrm{c})^6~\mathrm{MeV}^{-5}$]\\	
\midrule
(O): \ce{^3H},\ce{^4He}  \cite{LLE19} & -3.410 & -1.400\\
(I): \ce{^4He},\ce{^{24}Mg} & -3.00368 & -3.10429\\
(II): \ce{^3H},\ce{^4He},\ce{^{24}Mg} & -3.37705 & -1.78633\\
(III): \ce{^3H},\ce{^4He}, \ce{^{24}Mg} & -3.25392 & -2.20696\\
\bottomrule
\end{tabular}
\end{center}
\endgroup
\end{small}
\caption{Coupling constants of the adopted spin-isospin-symmetric interaction, along with the nuclei selected for the binding-energy-based constraints in the corresponding parameter adjustments. The values in (I), (II) and (III) are the fruit of the new fits presented in this work \cite{SEL26}. }\label{tab:coupling-constants}
\end{table*}

The configuration space of each nucleon is discretized and compactified into a cubic lattice of $N$ points per dimension, separated by a lattice spacing $a$. In the present calculations, $N = 10$ and the lattice spacing is fixed to $a \approx 1.316$~fm, thus the edges of the cube measure $L\equiv Na \approx 13.16$~fm. In turn, the time axis undergoes an analogous discretization into $N_t$ steps separated by a temporal lattice spacing of $a_t$. The latter is fixed to $a_t\approx~0.197~\mathrm{fm}/\mathrm{c}$ $\approx 10^{-3} \hbar~\mathrm{MeV}^{-1}$, while $N_t$ is varied up to $1000$ time steps in high-precision simulations. As a consequence, Schrödinger’s time evolution operator can be rewritten as a product of $N_t$ \textit{transfer matrices} \cite{LaM19,Lee09}.  

The transfer matrix is defined as the normal-ordered natural exponential of $-a_t H/\hbar$ and is expressed in terms of lattice second-quantization operators \cite{Lee09,Lee07}. The considered two and three-body spin-isospin-symmetric interactions contribute proportionally to the square and the cube of the nucleon-density operator, $\rho$, respectively. In the latter, creation and annihilation operators incorporate \textit{finite-size} effects through \textit{local} and \textit{non-local smearing} with respect to the nearest lattice sites \cite{SEL23}. For the dimensionless local and nonlocal smearing parameters \cite{LLE19}, the values $s_{\mathrm{L}}= 0.061$ and $s_{\mathrm{NL}}=0.5$ are adopted.

Eigenvalues and average values of physical observables are obtained through finite Euclidean-time evolution of trial states, $\Phi$. The expectation values of these quantities are represented by the infinite-Euclidean-time counterparts, which are extracted from samples of finite-$N_t$ observables through fitting functions of exponential type and real parameters, \textit{e.g.}
\begin{equation}
    E(L_t) = E_0 + \frac{c_0e^{-\Delta E_0 L_t/\hbar}}{1+c_0e^{-\Delta E_0 L_t/\hbar}}(E_0+\Delta E_0) \label{eqn:Energy-FitFunc-ETE}
\end{equation}
for energy eigenvalues, where $E_0$ is the $N_t\rightarrow +\infty$ limit, $\Delta E_0 >0$ the convergence rate and $c_0$ a constant. For further interpolating functions, one refers to the supplemental material of Ref.~\cite{SEL23}.

\subsection{Trial states}

The available \texttt{NLEFT} code, written in Fortran, \cite{SEL23} permits to construct trial wavefunctions of different nature as well as to perform multichannel calculations. The initial wavefunctions, $\Phi$, are recommended to have zero total momentum and sufficiently large overlap with the target eigenfunction, in the large Euclidean-time limit \cite{LaM19}. In order to select the total angular momentum, $J$, of the eigenfunctions in the continuum and infinite-volume limit, projectors \cite{Alt57,Joh82} to irreducible representations (irrep) of parity, $\mathscr{P}$, the cubic group, $\mathcal{O}$ \cite{LLL14,LLL15}, and the Abelian group $\mathcal{C}_4$ \cite{SEM18,Ste20}, are implemented. The latter two represent the lattice counterparts of the rotation groups SO(3) and SO(2) respectively. Decomposition tables for the irreducible representations of SO(3) with $J \leq 9$ into irreps of $\mathcal{O}$ are detailed in the appendix of Ref.~\cite{SSM22}.

Microscopic states of $\alpha$-cluster type are constructed by initializing the wavefunctions of the $A/4$ groupings of nucleons as spatially-distributed Gaussian wave-packets with width $w = 2.75~\mathrm{fm}$, \cite{LaM19, SLL21}, centered in selected lattice sites. So far, the arrangements in Fig.~1 of Ref.~\cite{SEL26} have been implemented, wherein the structure parameters are determined to reproduce the experimental value of the change radius of the nucleus, equal to $3.057(16)$~fm \cite{AnM13}. When multiple parameters are present, other observables of \ce{^{24}Mg} are reproduced, such as the position of the first minimum in the elastic form factor (bitetrahedron and staggered square bipyramid) or the electric quadrupole moment of the $2_1^+$ state \cite{StS26} (square bipyramid).

Furthermore, trial states of mean-field type consist of a single Slater determinant of $A$ single-particle 3D harmonic-oscillator (HO) wavefunctions in configuration space with frequency $\hbar\Omega = 41.0~A~\mathrm{MeV}$ by default, incorporating the spin and isospin degrees of freedom \cite{LaM19}.
Variants of such trial states have been developed by replacing the single-nucleon wavefunctions by linear combinations, with real coefficients, of HO eigenfunctions. 

As an ansatz, the coefficients have been extracted from the \texttt{KSHELL} \cite{SMU19} code. The programme enables large-scale nuclear shell-model calculations in $M$-scheme representation and suits the calculation of energy eigenvalues and eigenvectors, $M1$ and $E2$ moments, as well as transition probabilities. The final eigenvectors consist in superpositions of many $A$-body Slater determinants off HO wavefunctions, endowed with spin and isospin. Approximations of the \texttt{KSHELL} eigenvectors can be introduced by setting a lower cutoff in the overlap coefficients of the Slater-determinant expansion. In the \ce{^{24}Mg} case, it is convenient exploit USD interactions \cite{BrW88}, in combination with a valence space truncated to $8$ nucleons, corresponding to a \ce{^{16}O} core.

Although the mean-field states tested so far do not correspond with the latter, the admixtures in the single-particle HO wavefunctions represent an attempt to incorporate the complexity of \texttt{KSHELL} eigenvectors into single-Slater-determinant trial states. 
However, the implementation of mean-field trial states made of many-Slater determinants of single-nucleon HO eigenfunctions in the NLEFT code is envisaged \cite{SEL26}.

\section{Results}\label{sec:results}

For the present investigation on \ce{^{24}Mg}, three new fits for the parameters $C_2$ and $C_3$ have been performed, in order to fix the energy of the ground state of the nucleus on its experimental value of $-198.257040(2)$ MeV \cite{WHK21}, starting from the original set of parameters in the first row of Tab.~1 of Ref.~\cite{SEL26}. The first adjustment (I) is based on the measured energies of the $J^{\pi}=0^+$ ground states of \ce{^4He} and \ce{^{24}Mg}. Conversely, the second (II) hinges on the energy of the ground state of \ce{^{24}Mg}, and on the \textit{Tjon ratio} \cite{Tjo75}, \textit{i.e.} the ratio between the binding energies of \ce{^3H} and \ce{^4He}. The latter is experimentally equal to $\approx 0.299$ and represents as a manifestation of universality in nuclear systems \cite{PHM05,KEL18}.
Finally, the third (III) leverages the energy of the ground state of \ce{^{24}Mg}, minimizing at the same time the deviations between the ground state energies of \ce{^3H} and \ce{^4He} and their experimental counterparts.

Beneath $6$~MeV of excitation, the spectrum of \ce{^{24}Mg} consists of a $0_1^+$ state at -198.257 MeV, followed by a $2_1^+$ at -196.889 MeV, a $4_1^+$ at -194.134 MeV, a $2_2^+$ at -194.019 MeV and a $3_1^+$ at -193.022 MeV \cite{BaC22}. In the G$\alpha$CM with $\mathcal{D}_{4h}$ symmetry \cite{StS26}, the first three levels belong to the lowest $A_{1g}$ rotational band, whereas the last two are part of the lowest $B_{2g}$ band \cite{StS25,StS26}.

\subsection{Binding energies}

With the aim of testing the new coupling constants, the binding energy per nucleon of a series of self-conjugate-nuclei ranging from \ce{^{2}H} and \ce{^{40}Ca} has been calculated, considering a single value of $N_t$, fixed to $950$ time steps, which approximates rather well the result of an Euclidean-time extrapolation. The results are overall rather satisfactory, as the ones in Tab.~1 of Ref.~\cite{LLE19}. For the calculation of the binding energies, mean-field $A$-body states consisting of a single Slater determinant of HO wavefunctions have been adopted for the initialization of the trial states. For the nuclides with more marked open-shell character and $J^{\pi}\neq 0^{+}$, also single-nucleon-cluster trial wavefunctions \cite{LaM19} have been exploited. 

For the binding energies calculated with the original parameters (O), the sizable discrepancies between the binding energies in Ref.~\cite{LLE19} and the ones displayed in Fig.~2 of Ref.~\cite{SEL26} are due to the different lattice size adopted in the two cases, here fixed to $N =10$ for all nuclei in the sample. The parameters in the set (II) are capable of capturing the experimental trend of the binding energies well as reproducing simultaneously within $2$\% deviation the observed \textit{g.s.} energies \cite{WFL25} of $^{10}\mathrm{B}$, $^{18}\mathrm{F}$, $^{22}\mathrm{Na}$, $^{26}\mathrm{Al}$, $^{28}\mathrm{Si}$, $^{30}\mathrm{P}$ and $^{32}\mathrm{S}$ (cf. Fig.~2 of Ref.~\cite{SEL26}). The parameter set (III) delivers analogous outcomes for the latter nuclei, improving, in addition, the predictions for \ce{^{14}N} and \ce{^{22}Na}, at the price of displaying more sizable finite-volume effects for the three heaviest nuclei of the sample.  

\subsection{Ground state $0_1^+$}

With the parameter set (II), the energy of the $0_1^+$ state has been calculated starting from different $\alpha$-cluster and mean-field trial states by means of Euclidean-time extrapolations. Based on the latter, the parameter that gives a measure of the rapidness with which the energy eigenvalue at finite $L_t$ converges to the expectation value in the infinite Euclidean-time limit, $E_0$, coincides with $\Delta E_0$ in Eq.~\eqref{eqn:Energy-FitFunc-ETE}. The magnitude of $\Delta E_0$ reflects, in turn, the overlap between a trial state and the target eigenfunction of the considered Hamiltonian. Additionally, one constrains the interpolating exponentials in Eq.~\eqref{eqn:Energy-FitFunc-ETE} of energy samples associated with a different trial state to a common value of $E_0$, as in Figs.~2 and 4 of Ref.~\cite{SLL21}. This enables a more direct comparison between the convergence rates of the exponential curves, net of the value of $E_0$. 

For these calculations, the projector to the $A_1$ irreducible representation of the cubic group \cite{Car97}, corresponding to the $J=0$ irrep of SO(3) in the continuum and infinite-volume limit \cite{LLL14,LLL15,SEM18,SSM22}, has been exploited. Although unexploited, this projection operation can be skipped as long as the trial states possess a well-defined total angular momentum projection, $M$, along the $z$ axis \cite{SLL21}. 

\begin{figure}[htb!]
    \begin{center}
        \includegraphics[width=0.99\columnwidth]{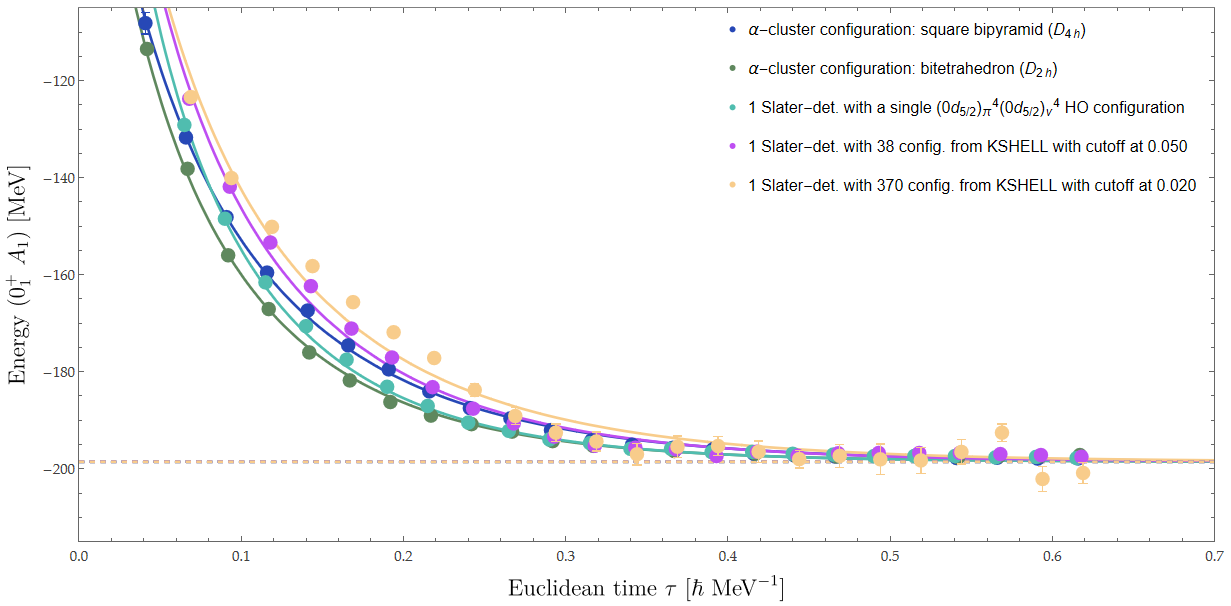}
    \end{center}
    \caption{\small{Transient energies of the $J^{\pi} =0^+$ ground state of \ce{^{24}Mg}, obtained from mean-field trial states (light yellow, purple and blue solid lines) as well as $\alpha$-cluster initial states, with $\mathcal{D}_{2h}$ and $\mathcal{D}_{4h}$ symmetry (acquamarina and green solid lines). The corresponding Euclidean-time extrapolated energy eigenvalue is denoted with dashed lines. For all samples, projectors to the $A_1^+$ irreps have been exploited.}}
      \label{fig:0+_I_Energy_AC-SM_ETE}
\end{figure}

Specifically, for the ground-state of \ce{^{24}Mg}, one finds that both the mean-field trial state with a single Slater-determinant of HO wavefunctions with the valence nucleons in the $(0d_{5/2})_{\pi}^4(0d_{5/2})_{\nu}^4$ configuration and the $\alpha$-cluster trial state with a bitetrahedral equilibrium structure have the largest convergence rate, $\Delta E_0 = 10.29(23)$ and $10.62(19)$ MeV respectively. The relevance of this triaxial configuration was first predicted by Ref.~\cite{HWD71}. Hence, it is legitimate to expect that the actual ground state of the nucleus possesses a dual nature, with preformed but overlapping $\alpha$-clusters. However, the prolate square bipyramidal configuration is associated with a rather high $\Delta E_0$ value, equal to $8.70(18)$ MeV.

Furthermore, a single-Slater determinant of made of a linear combination of $38$ HO eigenfunctions, whose coefficients correspond to the ones of a \texttt{KSHELL} eigenvector with $M=0$ and a cutoff of $0.050$ on the expansion coefficients (cf. Fig.~\ref{fig:0+_I_Energy_AC-SM_ETE}), does not improve the convergence rate of the mean-field trial state with single HO configurations, giving $\Delta E_0 = 9.06(16)$ MeV. The single-Slater-determinant trial state consisting of $370$ HO configurations, related to a \texttt{KSHELL} eigenvector with a cutoff of $0.020$ on the overlap coefficients, proves to have an even smaller convergence rate, $\Delta E_0 = 8.28(15)$ MeV. 

Beside the energy eigenvalue, both the elastic form factor and the charge density profile have been reconstructed by making use of the pinhole algorithm \cite{SEL23}. 
The results, detailed in Ref.~\cite{SEL26}, show that, starting from an $\mathcal{D}_{4h}$-symmetric  $\alpha$-cluster trial state, the position of the first minimum of the elastic form factor is shifted to a wavevector transfer of $\approx 1.65~\mathrm{fm}^{-1}$, instead of the measured value at $\approx 1.38~\mathrm{fm}^{-1}$. This fact is accompanied by the overestimation of the charge radius of the ground state. A pinhole calculation performed at $N_t = 916$, in fact, delivers $3.2974(199)$ fm, a value lying 7.9\% above the physical charge radius \cite{SEL26}.

\begin{figure}[htb!]
    \begin{center}
        \includegraphics[width=0.99\columnwidth]{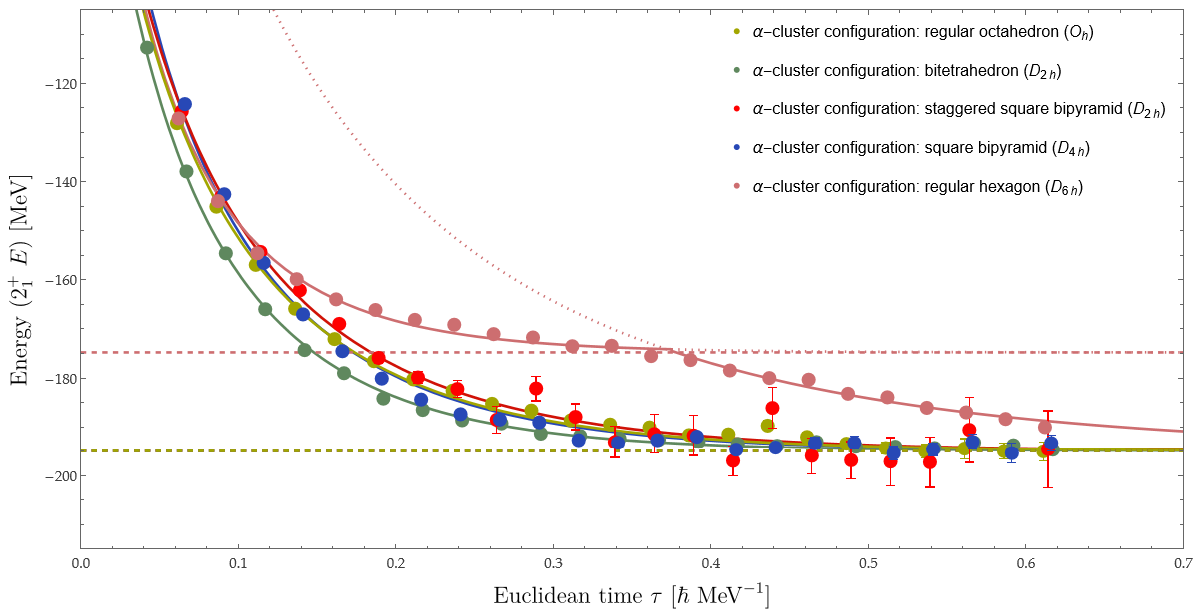}
    \end{center}
    \caption{\small{Transient energies of the $2_1^+$ excited state of \ce{^{24}Mg} with $E^+$ projectors, as a function of the Euclidean projection time $\tau$. The five datasets correspond to $\alpha$-cluster trial states with different point-group symmetry, but identical Gaussian wave-packet width, $w = 2.75$ fm. Solid lines denote the interpolating exponential functions. Dashed lines denote the infinite-Euclidean-time limit, $E_0$. For the sample corresponding to the hexagonal arrangement of $\alpha$-particles, the analytic continuation of the two interpolating curves outside their fitting domain is marked by dotted lines.}}
      \label{fig:2+_E_I_Energy_AC_ETE}
\end{figure}

\subsection{Excited state $2_1^+$}

For the lowest excitation of the \ce{^{24}Mg} nucleus, an even broader range of trial states has been exploited. Since angular momentum states with $J=2$ split into the $E\oplus T_2$ irreducible representations of the cubic group \cite{LLL14,LLL15,SEM18,SSM22}, the projectors to the $E$ or $T_2$ representations have been exploited, in combination with the one to positive parity \cite{SEL26}.  
Limiting ourselves to the $\alpha$-cluster trial states, all the configurations (a)-(e) in Fig.~1 of Ref.~\cite{SEL26} have been implemented for the simulations with $E$ projectors, obtaining an asymptotic energy eigenvalue of $-194.948(299)$~MeV. Analogously as for the $0_1^+$ state, the bitetrahedral structure appears to be most favoured in terms of convergence rate ($\Delta E_0 = 10.67(25)$~MeV), followed by the staggered square bipyramid ($9.39(19)$ MeV), the regular octahedron ($8.97(18)$~MeV) and the square bipyramid ($8.53(17)$~MeV). 

Conversely, the hexagonal configuration deserves a separate mention. Being oblate, such a configuration contrasts with the experimental electric quadrupole moment \cite{Fir07,KSG15}, hence unlikely to characterize the states belonging to the ground-state band \cite{BaC22}. The Euclidean-time extrapolation (cf. Fig.~\ref{fig:2+_E_I_Energy_AC_ETE}) seems to fully confirm this observation: the exponential function for the energies calculated at $N_t \leq 380$ seems to converge rapidly ($\Delta E_0 = 12.78(36)$~MeV) to an excited energy eigenvalue situated at $-175.363(643)$~MeV, compatible with the measured $(2^+)$ line at $20.83(2)$~MeV \cite{BaC22}. The latter might be the head of a superdeformed oblate rotational band. On the other hand, for $N_t > 380$, the energy data sample in Fig.~\ref{fig:2+_E_I_Energy_AC_ETE} follows a different exponential curve, slowly converging ($\Delta E_0 = 4.73(61)$~MeV) to the common value of $E_0$. This case underlines the importance of sampling the large-$N_t$ region, especially when the trial state is far from optimal. 

\section{Conclusion}

The low-lying $0_1^+$ and $2_1^+$ states of \ce{^{24}Mg} have been explored by means of a spin-isospin symmetric lattice Hamiltonian with smeared contact terms and Coulomb interaction. The new fits of the main coupling constants $C_2$ and $C_3$ turned out to provide a rather accurate description of the binding energies of self-conjugate \textit{sd}-shell nuclei, namely $^{10}\mathrm{B}$, $^{18}\mathrm{F}$, $^{22}\mathrm{Na}$, $^{26}\mathrm{Al}$, $^{28}\mathrm{Si}$, $^{30}\mathrm{P}$ and $^{32}\mathrm{S}$. Nonetheless, the $7.9$\% overestimation of the \ce{^{24}Mg} charge radius \cite{AnM13}, as well as the $0.27~\mathrm{fm}^{-1}$ shift in the position of the first minimum of the elastic form factor of the $0_1^+$ state, call for a new parameter adjustment, incorporating the local and nonlocal smearing parameters in the fit, with a the experimental charge radius as an additional constraint \cite{SEL26}. 

As emerged from the pilot studies for \ce{^{12}C} \cite{SEL23,SLL21}, the ground state band of \ce{^{24}Mg} seems to sit at a balance point, where both mean-field and $\alpha$-cluster effects contribute in shaping the eigenfunctions of the Hamiltonian. The values of the convergence rate of the energy eigenvalues in the Euclidean-time extrapolations suggest that, in the $0_1^+$ and $2_1^+$ states, the nucleus is likely to have a prolate shape, with sizable triaxiality. Among the $\alpha$-cluster trial states, the ones with smallest symmetry group, $\mathcal{D}_{2h}$, seem to be capable of accelerating the convergence process of the energy eigenvalues to the infinite $\tau$ limit. In fact, the bitetrahedral shape dominates in the $0_1^+$ ($A_1$) and $2_1^+$($E$) states, whereas for the $2_1^+$($T_2$) eigenstate it is challenged by the staggered squared bipyramid \cite{SEL26}. 
In turn, the square bipyramidal configuration is favoured over the octahedral one, whereas the hexagonal structure seems to play a role only at excitation energy~$\gtrsim 20$ MeV. 

Eventually, only the planned reconstruction of the nuclear density profile as in Ref.~\cite{SEL23} will permit to unravel the arrangement of protons and neutrons into $\alpha$ particles in the final $0_1^+$ and $2_1^+$ eigenvectors \cite{SEL26}.

\section*{Acknowledgements}

G.S. expresses gratitude to Dean Lee (MSU, East Lansing, USA), Bing-Nan Lu (GSCAEP, Beihang University, China) and Teng Wang (University of Beijing, China) for the fruitful discussions and acknowledges financial support from the CNRS/IJCLab/In2p3, unit UMR9012.
For the realization of this work, we acknowlegde computational resources provided by the \textit{Jean Zay} cluster of the \textit{Idris} facility of the \textit{Grand équipement national de calcul intensif} (GENCI) and the \textit{Irène} cluster of the \textit{Très Grand Centre de Calcul} (TGCC), in the framework of the eDARI project No. A010516546.


\begin{thebibliography}{99}
\bibitem{Wef37} 
W. Wefelmeier, \href{https://link.springer.com/article/10.1007/BF01330174}{\emph{Zeitschr. für Physik} \textbf{107}, 332 (1937)}.
\bibitem{Whe37} 
J.A. Wheeler, \href{https://journals.aps.org/pr/abstract/10.1103/PhysRev.52.1083}{\emph{Phys. Rev.} \textbf{52}, 1083 (1937)}.
\bibitem{KaE19} 
Y. Kanada-En'yo, "\textit{Clustering in light neutron-rich nuclei}" in P. Van Duppen, A. Vitturi and S. Pirrone, \emph{Proc. Int. Sch. Phys. "Enrico Fermi"} \textbf{201}, 61-93, Varenna (2019). 
\bibitem{Ohk22} 
S. Ohkubo, \href{https://journals.aps.org/prc/abstract/10.1103/PhysRevC.106.034324}{\emph{Phys. Rev. C} \textbf{106}, 034324 (2022)}.
\bibitem{SEL23} 
S. Shen, S. Elhatisari, T.A. L\"ahde, D. Lee, B.-N. Lu and U-.G. Mei\ss{}ner, \href{https://www.nature.com/articles/s41467-023-38391-y}{\emph{Nature Comm.} \textbf{14}, 2777 (2023)}.
\bibitem{Nes23} 
V.O. Nesterenko, \href{https://iopscience.iop.org/article/10.1088/1742-6596/2586/1/012074}{\emph{J. of Phys. Conf. Ser.} \textbf{2586}, 012074 (2023)}.
\bibitem{CRD23} 
J. Cseh, G. Riczu and J. Darai, \href{https://www.mdpi.com/2073-8994/15/12/2116}{\emph{Symmetry} \textbf{15}, 115 (2023)}.
\bibitem{HoB24} 
H. Horiuchi and D. Blaschke, \href{https://link.springer.com/article/10.1140/epja/s10050-024-01319-1}{\emph{Eur. Phys. J. A} \textbf{60}, 197 (2024)}.
\bibitem{AZZ26} 
D.A. Artemenkov, A.A. Zaitsev and P.I. Zarubin, \href{https://nsr.jinr.int/article/92}{\emph{Nat. Sci. Rev.} \textbf{3}. 200603 (2026)}.
\bibitem{Gam30} 
G. Gamow, \href{https://royalsocietypublishing.org/rspa/article/126/803/632/2732/Mass-defect-curve-and-nuclear-constitution}{"\emph{Mass-defect curve and nuclear constitution}", in E. Rutherford, \emph{R. Soc. Lond. Proc. Ser. A} \textbf{126}, 632-644 (1930)}.
\bibitem{HaT38} 
L.R. Hafstad and E. Teller, \href{https://journals.aps.org/pr/abstract/10.1103/PhysRev.54.681}{\emph{Phys. Rev.} \textbf{54}, 681 (1938)}.
\bibitem{Car97} 
R.L. Carter, \textit{Molecular Symmetry and Group Theory}, John Wiley \& Sons (1997).
\bibitem{DeD24} 
I. Dedes and J. Dudek, "\emph{$\mathcal{C}_{2v}$ (Water Molecule) Symmetry Identified in an Actinide Nucleus \ce{^{236}U}}", SSNET 2024 Conference, Orsay, France (presentation).
\bibitem{BKB25} 
S. Basak, D. Kumar, T. Bhattacharjee, I. Dedes, J. Dudek, A. Pal, S.S. Alam, A. Saha, A.K. Sikdar et al., \href{https://journals.aps.org/prc/abstract/10.1103/PhysRevC.111.034319}{\emph{Phys. Rev. C} \textbf{111}, 034319 (2025)}.
\bibitem{BiI02} 
R. Bijker and F. Iachello, 
\href{https://www.sciencedirect.com/science/article/pii/S000349160296255X}{\emph{Ann. of Phys.} \textbf{298},2, 334 (2002)}.
\bibitem{KiT24} 
M. Kimura and Y. Taniguchi, \href{https://link.springer.com/article/10.1140/epja/s10050-024-01302-w}{\emph{Eur. Phys. J. A} \textbf{60}, 77, (2024)}.
\bibitem{StS25} 
G. Stellin and K.-H. Speidel, "\emph{Spectrum and electromagnetic properties of $^{24}\mathrm{Mg}$ in the Geometric $\alpha$-cluster Model with $\mathcal{D}_{4h}$ symmetry at leading order}" in M. Gaidarov and N. Minkov, \emph{Nuclear Theory} \textbf{42}, Heron Press, Sofia (2025).
\bibitem{StS26} 
G. Stellin and K.-H. Speidel, \href{https://iopscience.iop.org/article/10.1088/1361-6471/ae384c}{\emph{J. of Phys. G.} \textbf{53}, 025102 (2026)}.
\bibitem{Bou62} 
M. Bouten,
\href{https://link.springer.com/article/10.1007/BF02754343}{\emph{Nuovo Cimento} \textbf{26}, 1, 3895-3904, (1962)}.
\bibitem{HaD66} 
P.S. Hauge and G.H. Duffey, \href{https://journals.aps.org/pr/abstract/10.1103/PhysRev.152.1023}{\emph{Phys. Rev.} \textbf{152}, 1023 (1966)}.
\bibitem{Ste15} 
G. Stellin, \href{http://tesi.cab.unipd.it/50308/}{\emph{Simmetrie e rotovibrazioni di nuclei $\alpha$-coniugati}, M.Sc. Thesis, Dipartimento di Fisica e Astronomia "G. Galilei", Università degli Studi di Padova (2015)}.
\bibitem{SFV16} 
G. Stellin, L. Fortunato and A. Vitturi, \href{https://iopscience.iop.org/article/10.1088/0954-3899/43/8/085104}{\emph{J. Phys. G} \textbf{43}, 085104 (2016)}.
\bibitem{HWD71} 
P.S. Hauge, S.A. Williams and G.H. Duffey, \href{https://link.aps.org/doi/10.1103/PhysRevC.4.1044}{\textit{Phys. Rev. C} \textbf{4}, 4, 1044-1061 (1971)}.
\bibitem{EKL12} 
E. Epelbaum, H. Krebs, T.A. L\"ahde, D. Lee and U.-G. Mei\ss{}ner, \href{https://doi.org/10.1103/PhysRevLett.109.252501}{\emph{Phys. Rev. Lett.} \textbf{109}, 252501 (2012)}.
\bibitem{EKL14} 
E. Epelbaum, H. Krebs, T.A. L\"ahde, D. Lee, U.-G. Mei\ss{}ner and G. Rupak, \href{https://doi.org/10.1103/PhysRevLett.112.102501}{\emph{Phys. Rev. Lett.} \textbf{112}, 10, 102501 (2014)}.
\bibitem{KBL18} 
C. K\"orber, E. Berkowitz et T. Luu, \href{https://doi.org/10.1051/epjconf/201817511012}{\emph{Eur. Phys. J. Web of Conf.} \textbf{175}, 11012 (2018)}.
\bibitem{EBM24} 
 S. Elhatisari, L. Bovermann, Y.-Z. Ma, E. Epelbaum, D. Frame, F. Hildenbrand, M. Kim, Y. Kim, H. Krebs, T. A. L\"ahde, D. Lee, N. Li, B.-N. Lu, U.-G. Meißner, G. Rupak, S. Shen, Y.-H. Song and G. Stellin, \href{https://www.nature.com/articles/s41586-024-07422-z}{\emph{Nature} \textbf{630}, 59–63 (2024)}.
 \bibitem{LaM19} 
T.A. L\"ahde et U.-G. Mei\ss{}ner, \href{https://doi.org/10.1007/978-3-030-14189-9}{\emph{"Nuclear Lattice Effective Field Theory: An Introduction"}, \emph{Lecture Notes in Physics} \textbf{975}, Springer (2019)}.
\bibitem{LLE19} 
B.-N. Lu, N. Li, S. Elhatisari, D. Lee, E. Epelbaum and U.-G. Mei\ss{}ner, \href{http://doi.org/10.1016/j.physletb.2019.134863}{\emph{Phys. Lett. B} \textbf{797}, 134863 (2019)}. 
\bibitem{SEL26} 
G. Stellin, S. Elhatisari, T.A. L\"ahde and S. Shen, "\textit{Wigner SU(4) symmetry, clustering, and the spectrum of \ce{^{24}Mg}}" (unpublished).
\bibitem{YKL23} 
H. Yu, S. K\"onig and D. Lee, \href{https://journals.aps.org/prl/abstract/10.1103/PhysRevLett.131.212502}{\emph{Phys. Rev. Lett.} \textbf{131}, 212502 (2023)}.
\bibitem{Lee09} 
 D. Lee,
\href{https://www.sciencedirect.com/science/article/abs/pii/S014664100800094X}{\emph{Progr. in Part. and Nucl. Phys.} \emph{63}, 117-154 (2009)}.
\bibitem{MaS16} 
R. Machleidt and F. Sammarruca, \href{https://iopscience.iop.org/article/10.1088/0031-8949/91/8/083007}{\emph{Phys. Scripta} \textbf{91}, 083007 (2016)}.
\bibitem{Heb21} 
K. Hebeler, \href{https://www.sciencedirect.com/science/article/abs/pii/S0370157320303409?via%3Dihub}{\emph{Phys. Rep.} \textbf{890}, 1-116 (2021)}.
\bibitem{MRR53} 
N. Metropolis, A.W. Rosenbluth, M.N. Rosenbluth and A.H. Teller, \href{https://doi.org/10.1063/1.1699114}{\emph{J. Chem. Phys.} \textbf{21}, 1087-1092 (1953)}.
\bibitem{Lee07} 
D. Lee, \href{https://doi.org/10.1103/PhysRevLett.98.182501}{\emph{Phys. Rev. Lett.} \textbf{98}, 182501 (2007)}.
\bibitem{Alt57} 
S. Altmann, \href{https://www.cambridge.org/core/journals/mathematical-proceedings-of-the-cambridge-philosophical-society/article/on-the-symmetries-of-spherical-harmonics/37E30295C053B36516BBFBDDED423826}{\emph{Math. Proc. Phyl. Soc.} \textbf{53}, 2, 343-367 (1957)}.
\bibitem{Joh82} 
R.C. Johnson, \href{https://www.sciencedirect.com/science/article/pii/0370269382901344}{\emph{Phys. Lett.} \textbf{114 B}, 147-151 (1982)}.
\bibitem{LLL14} 
Bing-Nan Lu, Timo A. L\"ahde, Dean Lee, and Ulf-G. Mei\ss{}ner, \href{https://journals.aps.org/prd/abstract/10.1103/PhysRevD.90.034507}{\emph{Phys. Rev. D} \textbf{90}, 034507 (2014)}.
\bibitem{LLL15} 
Bing-Nan Lu, Timo A. L\"ahde, Dean Lee, and Ulf-G. Mei\ss{}ner, \href{https://journals.aps.org/prd/abstract/10.1103/PhysRevD.92.014506}{\emph{Phys. Rev. D} \textbf{92}, 014506 (2015)}.
\bibitem{SEM18} 
G. Stellin, S. Elhatisari and U.-G. Meißner, \href{https://epja.epj.org/articles/epja/abs/2018/12/10050_2018_Article_12671/10050_2018_Article_12671.html}{\emph{Eur. Phys. J. A} \textbf{54}, 232 (2018)}. 
\bibitem{Ste20} 
G. Stellin, \href{https://bonndoc.ulb.uni-bonn.de/xmlui/handle/20.500.11811/8892}{\textit{Nuclear Physics in a finite volume: Investigation of two-particle and $\alpha$-cluster systems}, Ph.D. Thesis, HISKP, Universität Bonn, (2020)}.
\bibitem{SSM22} 
G. Stellin, K.-H. Speidel and U.-G. Mei\ss{}ner, \href{https://epja.epj.org/articles/epja/abs/2022/10/10050_2022_Article_850/10050_2022_Article_850.html}{\textit{Eur. Phys. J. A} \textbf{58}, 208 (2022)}.
\bibitem{SLL21} 
S. Shen, T.A. L\"ahde, D. Lee and U.-G. Meißner, \href{https://epja.epj.org/articles/epja/abs/2021/09/10050_2021_Article_586/10050_2021_Article_586.html}{\emph{Eur. Phys. J. A} \textbf{57}, 276 (2021)}.
\bibitem{AnM13} 
I. Angeli and A. Marinova, \href{https://www.sciencedirect.com/science/article/abs/pii/S0092640X12000265}{\emph{Atom. Data and Nucl. Data Tab.} \textbf{99}, 1, 69-95 (2013)}.
\bibitem{SMU19} 
N. Shimizu, T. Mizusaki, Y. Utsuno and Y. Tsunoda, \href{https://www.sciencedirect.com/science/article/abs/pii/S0010465519301985}{\emph{Comp. Phys. Comm.} \textbf{244}, 372-384 (2019)}.
\bibitem{WHK21} 
M. Wang, W.J. Huang, F.G. Kondev, G. Audi and S. Naimi, \href{https://doi.org/10.1088/1674-1137/abddaf}{\emph{Chin. Phys. C} \textbf{45}, 030003 (2021)}.
\bibitem{Tjo75} 
J.A. Tjon, \href{https://www.sciencedirect.com/science/article/abs/pii/0370269375903780}{\emph{Phys. Lett. B} \textbf{56}, 3, 217-220 (1975)}.
\bibitem{PHM05} 
L. Platter, H.-W. Hammer and U.-G. Mei\ss{}ner, \href{https://doi.org/10.1016/j.physletb.2004.12.068}{Phys. Lett. B 607, 254 (2005)}.
\bibitem{KEL18} 
N. Klein, S. Elhatisari, T.A. L\"ahde, D. Lee, U.-G. Mei\ss{}ner, \href{https://epja.epj.org/articles/epja/abs/2018/07/10050_2018_Article_12553/10050_2018_Article_12553.html}{\emph{Eur. Phys. J. A} \textbf{54}, 121 (2018)}.
\bibitem{BaC22} 
M.S. Basunia and A. Chakraborty, \href{https://www.sciencedirect.com/science/article/abs/pii/S0090375222000576}{\emph{Nucl. Data Sheets} \textbf{186}, 3-262 (2022)}.
\bibitem{WFL25} 
T. Wang, X. Feng and B.-N. Lu, \href{https://arxiv.org/abs/2512.21942}{\emph{ArXiv} 2512.21942, (2025)}.
\bibitem{BrW88} 
B.A. Brown and B.H. Wildenthal, \href{https://www.annualreviews.org/content/journals/10.1146/annurev.ns.38.120188.000333}{\textit{Annu. Rev. Nucl. Part. Sci.} \textbf{38}, 29 (1988)}.
\bibitem{Fir07} 
R.B. Firestone, \href{https://doi.org/10.1016/j.nds.2007.10.001}{\emph{Nucl. Data Sheets} \textbf{108}, 11, 2319-2392 (2007)}.
\bibitem{KSG15} 
A. Kusoglu, A. E. Stuchbery, G. Georgiev, B.A. Brown, A. Goasduff, L. Atanasova, D.L. Balabanski, M. Bostan, M. Danchev, P. Detistov, K.A. Gladnishki, J. Ljungvall, I. Matea, D. Radeck, C. Sotty, I. Stefan, D. Verney and D.T. Yordanov, \href{https://doi.org/10.1103/PhysRevLett.114.062501}{\emph{Phys. Rev. Lett.} \textbf{114}, 6, 062501 (2015)}. 

\end{thebibliography}
\end{document}